\documentstyle[twoside,fleqn,espcrc2]{article}

\newcommand{\LQ}{\Lambda_{QCD}}
\newcommand{\be}{\begin{equation}}
\newcommand{\ee}{\end{equation}}
\newcommand{\beqn}{\begin{eqnarray}}
\newcommand{\eeqn}{\end{eqnarray}}
\newcommand{\bea}{\begin{eqnarray}}
\newcommand{\eea}{\end{eqnarray}}
\newcommand{\La}{\Lambda_{QCD}}
\newcommand{\beaq}[1]{\begin{equation}\begin{array}{#1}}
\newcommand{\eeaq}{\end{array}\end{equation}}
\newcommand{\D}[1]{{\displaystyle #1}}

\title{Power corrections from small distances.
\thanks{Talk presented by V.I. Zakharov at the Euroconference on Quantum
Chromodynamics, QCD'99, Montpellier, July 1999.}}
\author{F.V. Gubarev, M.I. Polikarpov,\address{
Institute of Theoretical and Experimental Physics, \\
       B. Cheremushkinskaya, 25, 
117259 Moscow }%
    and
        V.I. Zakharov\address{Max-Planck Institut f\"ur Physik, \\
F\"ohringer Ring 6, 80805 M\"unchen, Germany.
        }}

\begin{document}

\begin{abstract}We review recent speculations on
power like corrections in QCD which go beyond the
standard Operator Product Expansion. Both the theoretical picture 
underlying
these corrections and phenomenological manifestations are 
discussed in some
detail. 
\end{abstract}

\maketitle
\section{Introduction}

Power corrections have been discussed intensely
in recent years, in particular,
in review talks at the conferences in this series \cite{avz}.
Thus, we would skip the general motivation 
and background. Instead, we will concentrate on
one particular issue, that is speculations on 
hypothetical power corrections
associated with short distances \cite{gpz,cnz}.
For the talk to be self contained, however, we will include also a very
brief review of the standard picture
based on the operator product expansion and
underlying the QCD sum rules \cite{svz}. 
Naturally enough, in the standard picture we
emphasize only the points 
which would be modified if there exist novel power corrections.

Thus, the outline of the talk as follows:

1. Standard picture.\\
2. Standard picture vs experiment.\\
3. Beyond the OPE.\\
4. Predictions, tests.

\section{Standard picture.}

Many of the ideas belonging now to the standard picture could in fact
be traced back to a paper which is 51 years old, that is the paper
by Casimir and Polder \cite{casimir}. It is a pleasure to quote this
paper which in fact is much more elaborated than
the quotations may indicate. 
The first example of what we would interpret nowadays as a power
correction can be understood by everybody.
Namely, consider an $e^+e^-$ pair at distance $r$ placed into
a center of a conducting cage of size $L$. Moreover, assume
that $L\gg r$. Then the potential energy of the pair can be 
approximated as 
\be
V_{e\bar{e}}(r)\approx-{\alpha_e\over r}+(const){\alpha_e r^2\over L^3},
~~~L\gg r\label{cage}
\ee
and the second term is a power correction to the Coulomb interaction.
The derivation of (\ref{cage}) is of course straightforward in
terms of the classical electrodynamics, since the correction is
nothing else but the interaction of the dipole with its images.
On the other hand, it can be derived also in terms of one-photon
exchange. Moreover, to find the photon propagator one should find now the modes
in the cage which are different from those in the empty space.
The difference is of order unity at frequencies of order 
$\omega\sim 1/L$.

Now, we jump to QCD and conclude by analogy that the heavy quark
potential in QCD looks at short distances as:
\be
\lim_{r\to 0}V(r)~=~-{c_{-1}\over r}+const \cdot \LQ^3r^2,\label{potential}
\ee
where $c_{-1}$ is calculable perturbatively as a series in $\alpha_s$.
Eq. (\ref{potential})
was derived first in Ref. \cite{balitsky}. 
Note the absence of a linear correction to the potential at 
short distances.
The logic behind (\ref{potential}) is that
we simply replace 
$L\to \LQ^{-1}$ since the gluon propagator is modified by
the infrared effects at $\omega\sim \La^{-1}$. 
 
If one turns to consideration of bound states, then the quadratic 
correction
in the potential (\ref{cage}) is washed out by the retardation effects
\cite{casimir}. Indeed, if $T$ is the period of rotation of 
the charges and 
$c$ is the speed of light then there might be not time enough to learn
about the existence of the cage. Thus, if
\be
T\cdot c~ \ll ~ L,
\ee
then the retardation effects are crucial and one cannot use the potential
(\ref{cage}). Instead, the shifts of the atomic levels in the cage
are sensitive to a local characteristic of the non-perturbative fields
which can be nothing else but $<0|{\bf E}^2|0>_{non-pert}$. 
On dimensional grounds:
 \be
(\delta E)_{non-pert}~\sim {<0|{\bf E}^2|0>_{non-pert} \over m_e^3}\label{qed}
,\ee
where by $<0|{\bf E}^2|0>_{non-pert}$ one understands the difference between
the average valued of ${\bf E}^2$ in one-photon approximation
evaluated without and with the cage. Note that all the ultraviolet
divergences cancel in this difference.

In fact $<0|{\bf E}^2|0>_{non-pert}$ was not introduced in \cite{casimir}
and only the numerical result was given which is possible
since $<0|{\bf E}^2|0>_{non-pert}\sim L^{-4}$, with all the
coefficients calculable. 
Thus, we brought Eq. (\ref{qed}) to 
a the modern form which first appeared 
within the Voloshin-Leutwyler picture for $Q\bar{Q}$
bound states \cite{voloshin}.
In that case the corresponding density of the color field strengths, 
or the
gluonic condensate $<0|\alpha_s(G^a_{\mu\nu})^2|0>$,
cannot be calculated directly
but can be extracted from independent data
\cite{svz}.
Thus, an analog of (\ref{qed}) looks as
\be
\delta E_{n,l}~=~f_{n,l}{<0|\alpha_s(G^a_{\mu\nu})^2|0> \over m_Q^4}
+(pert.~th.)\label{levels}
\ee
where $f_{n, l}$ encodes the dependence on the quantum numbers
of the $Q\bar{Q}$ bound state and we included all the constants, 
such as powers of
$\alpha_s$, into the $f_{n,l}$. Further details and references can be
found, e.g., in the review \cite{yndurain}. 

Another important prediction of the standard model
is the absence of 
$1/Q^2$ corrections to the current correlation functions $\Pi_j(Q^2)$
at large $Q^2$~\cite{svz}. The correlation functions are defined as
\be
\Pi_j(Q^2)~=~i\int e^{iqx}d^4x\langle 0|T\{j(x),j(0)\}|0\rangle,
\ee
where $q^2\equiv -Q^2$ and
$j(x)$ are local currents constructed on the quark and gluon fields
and we suppress possible Lorenz indices.

Then, basing on the OPE, one can write at large $Q^2$:
\beaq{c}
\Pi_j(Q^2)\approx~ \Pi_j(Q^2)_{parton~model}\cdot
\\
\\
\left[ 1+{\D{a_j}\over \D{\ln[Q^2/\LQ^2]}}+
{\D{b_j}\over \D{Q^4}}+ O((lnQ^2)^{-2},Q^{-6})\right]
\label{normal}\eeaq
where the constants $a_j,b_j$ depend on the channel, i.e. on quantum
numbers of the current $j$. Terms of order $1/lnQ^2$ and $Q^{-4}$
are representing the first perturbative correction and the gluon
condensate, respectively. 
The absence of the linear correction in $1/Q^2$ is an analog
of the absence of the linear correction to the potential at short
distances (\ref{potential}).

{\it To summarize,} the standard picture assumes that the only change
brought
by non-perturbative fluctuations is the
change in the ordinary Feynman graphs  
at virtual momenta of order $\LQ^{-1}$. 
The power corrections parameterize these changes. Among the predictions
of the standard picture are the absence of linear corrections
to the $Q\bar{Q}$ potential at short distances $V(r)$ 
and of $1/Q^2$ corrections
to the current correlation functions $\Pi_j(Q^2)$.

\section{Standard picture vs experiment.}

We are going now to quote a few results which, taken at face value, would
rule the standard picture out. From the beginning, however, we would like
to avoid any dramatic statement. It is all the more so that the
authors of the papers quoted do not draw such conclusions themselves.
In each case there could be specific problems, like subtraction of
the perturbative contributions and we are not in position at all to
analyze the data thoroughly. We would simply like to attract attention
to the kind of measurements 
which can be crucial to check the standard picture.

(1){\it The heavy quark potential at small $r$.}

As is mentioned above the standard picture predicts the absence
of the linear correction at $r\to 0$, see (\ref{potential}).
The existing data are rather fitted to the form:
\be
\lim_{r\to 0}{V_{Q\bar{Q}}(r)}~=~-{c_{-1}\over r}+\sigma_0\cdot r \label{sigma0}
.\ee
As for the numerical value of $\sigma_0$ it is conveniently expressed
in units of $\sigma_{\infty}$ where $\sigma_{\infty}$ which is the
string tension at large distances. Then:
\be
\sigma_0~=~(1~\div~5)\sigma_{\infty}.
\ee
Here the value $\sigma_0\approx\sigma_{\infty}$ can be extracted 
as an estimate from the existing data for all $r$ \cite{schilling}
while $\sigma_-\approx 5\sigma_{\infty}$ is the result of a very recent
dedicated study of small distances \cite{bali}. The number 
is sensitive to the subtraction of the perturbative terms,
see \cite{peter} and the Talk at this conference by Y. Schroeder.

A new and exciting perspective to measure the non-perturbative
correction to the potential is provided by the so called
P-vortices (for latest results and further references see, e.g.,
\cite{bertle}). The P-vortices are particular non-perturbative 
field configurations observed on the lattice. The property 
which is crucial
for us at the moment is that they dominate non-perturbative contributions
and, in particular, reproduce the string tension $\sigma_{\infty}$
as defined in terms of the static potential $V(r)$ at $r\to \infty$.
Moreover, if one leaves the contribution of the P-vortices alone, 
then the Coulombic part is removed from $V(r)$ altogether. 
The non-perturbative potential $V(r)_{non-pert}$ defined now as arising
from the P-vortices alone shows no change in the linear
behavior down to $r$ of order of the lattice size $a$ \cite{deldebbio}:
\be
\lim_{r\to 0}{V(r)_{non-pert}}~\approx~\sigma_{\infty}\cdot r.
\ee

Finally, analytical studies
of the fine structure of the charmonium levels
also prefer strongly $\sigma_0\approx\sigma_{\infty}$ \cite{badalian}.

(2) {\it Shifts of the $Q\bar{Q}$ atomic levels.}

On the lattice, one can measure the energy levels of the heavy $Q\bar{Q}$
systems as a function of the quark mass \footnote{We are thankful to 
A. Leonidov for bringing our attention to this kind of measurements.}.
The latest measurements of this kind can be found in Ref. \cite{fingberg}.
Using Eq. (\ref{levels}) one can 
interpret the results of the measurements
in terms of the gluon condensate:
\be
<0|\alpha_s(G^a_{\mu\nu})^2|0>_{exp}~=~f(m_Q)
\ee
where the standard picture corresponds to the function $f(m_Q)=const$.
In reality, as the quark mass $m_Q$ varies between $\sim 5$ and $\sim$50 GeV
the function $f(m_Q)$ varies by a factor $15\div 20$ \cite{fingberg}.\\

(3) {\it Correlation functions.}

It has been known since long \cite{novikov} that at least in some, so
to say exceptional channels the sum rules (\ref{normal}) fail.
Namely, one characterizes different channels by $M^2_{crit}$
which is defined as the Euclidean mass scale where the asymptotic freedom
is violated by the power corrections by 10\%. Then the values
of $M^2_{crit}$ vary considerably in some cases:
\beqn
(M^2_{crit})_{\rho-meson~}\approx~ 0.6 \; GeV^2,\label{rho}\\ 
(M^2_{crit})_{\pi-meson}~\approx~ 2 \; GeV^2,\label{pi}\\
(M^2_{crit})_{0^+glueball}~\approx~ 20 \; GeV^2.\label{0} 
\eeqn
The estimates were obtained first by indirect means
using theoretical input specific for the particular channel \cite{novikov}.
Moreover, the difference between the $\rho-$ and $\pi-$ channels is fully 
confirmed by the lattice measurements, see Ref. \cite{chu}. 
The standard picture, see Eq. (\ref{normal}), fails completely to
reproduce the hierarchy of the mass scales.
There are fits to the data based on parameterization of instanton
contributions \cite{shuryak,novikov}. The model works well,
with a notable exception of the $\sigma$-meson channel, see \cite{chu}.

{\it To summarize}, the standard picture seems to disagree with a number
of measurements. The best known case is the correlation functions
in some channels (see point (3) above). In this case there were attempts
to amend the situation by inclusion of the direct instantons.
Measurements of the potential at short distances 
and the $Q\bar{Q}$ energy levels (points (1) and (2) above) are quite 
recent and have not been interpreted much.

\section{Beyond the OPE. Compact photodynamics.}

Soft corners of the phase space caught by the OPE is certainly
not the only source of power corrections. As we will emphasize in this 
and next sections topological excitations could be also hidden 
behind the power
corrections \cite{gpz}. Whether this mechanism applies to QCD
is an open question however.

The first example of a theory where the OPE does not work 
is very old in fact and goes back to the paper in Ref. \cite{polyakov}.
However, 
since it has not been discussed in connection with the OPE, we will
explain this example in some detail.

The action of the theory we are going to consider is very simple:
\be
S~=~{1\over 4 e^2}\int d^4x F^2_{\mu\nu}\label{action}
\ee
where $F_{\mu\nu}$ is the Abelian field strength tensor, 
$F_{\mu\nu}=\partial_{\mu}A_{\nu}-\partial_{\nu}A_{\mu}$.
The action (\ref{action}) is that of free photons
and at first sight nothing interesting can come out from
this theory. 
In particular, if we introduce external static
electric charges as a probe, their potential energy would be given by
one-photon exchange without any corrections.

However, we shall see in a moment that, in a particular formulation,
the theory admits also magnetic monopoles. Hence,
a few preliminary words on the monopoles.
Monopoles have magnetic field similar to the electric field of a charge:
\be
{\bf H}~=~q_M {{\bf r}\over 4\pi r^3}
.\ee
Then the flux of the magnetic field through a surface surrounding
the monopole is:
\be
\Phi~=~q_M
.\ee
On the other hand, because of the equation $div~{\bf H}=0$ 
the magnetic flux is conserved. 
Thus, the magnetic monopole cannot exist by itself
and one assumes that there is a string
which is connected to the monopole and which brings in the flux.
Moreover, to make the string invisible
one assumes that the string is infinitely thin. Finally, to
avoid the Bohm-Aharonov effect
one imposes the Dirac quantization condition,
\be
e\oint A_{\mu}dx_{\mu}~=~e\int {\bf H}\cdot d{\bf s}~=q_M\cdot e
~=~2\pi\cdot n
\ee
Also, we ask for the energy (or action) associated with the  
Dirac string to vanish. Only then energies of the electric and 
magnetic charges 
are similar. We shall return to discuss the issue of the
energy of the Dirac string later. 

Now, the Dirac strings may end up with monopoles. The action
associated with the monopoles is not zero at al but rather
diverges in ultraviolet, since
\be 
\int {d^3r\over 8\pi}{\bf H}^2~\sim~{1\over e^2a}\label{uvdivergence}
\ee
where $a$ is a (small) spatial cut off. If the length
of a closed monopole trajectory is L, then the
suppression of such a configuration due to
a non-vanishing action is of order
\be
e^{-S}~\sim~\exp~(-const~L/e^2)
.\ee
On the other hand, there are different ways to organize a loop of
length $L$. This is 
the entropy factor. It is known to grow exponentially with $L$ as
$\sim \exp(~const'L~)$.

Thus, one comes to the conclusion 
\cite{polyakov} that at some $e_{crit}\sim 1$ there occurs a phase 
transition corresponding to condensation of the monopole loops.
As a result, if external electric charges are introduced as a probe,
their potential energy grows with distance, $V(r)\sim r$ and
they are confined. 

To complete the presentation we should explain how one should
understand the theory (\ref{action}) that it would imply
a vanishing action for the Dirac string.

The crucial point is to define the theory
by means of a lattice regularization.
Then the action can be understood as a sum over plaquette
actions:
\beqn
S~=\sum{1\over 2e^2}\left(1-~Re~ \exp(i\oint A_{\mu}dx^{\mu}
\right)~=~
\nonumber \\
~=~\sum{1\over 2 e^2}\left(1-Re~ \exp(i F_{\mu\nu}d\sigma_{\mu\nu})\right)~=~\\
~=~\sum{1\over 2 e^2}\left(1-\cos(F_{\mu\nu}d\sigma_{\mu\nu})\right)
\nonumber 
\eeqn
where the sum is taken over all plaquettes (and note 
that no summation over the 
repeated indices $\mu,\nu$ is understood).
In the continuum limit one reproduces of course 
the action (\ref{action}). However, from the intermediate steps it 
is clear
that the action admits for a large jump in $F_{\mu\nu}$:
\be
F_{\mu\nu}~\rightarrow~F_{\mu\nu}+2\pi\delta (\sigma_{\mu\nu})\label{dstring}
\ee
where the $\delta$-function on the surface is defined
as $\delta(\sigma_{\mu\nu})d\sigma^{\mu\nu} =1$ (no summation over $\mu,\nu$).
The second term in Eq. (\ref{dstring}) exactly corresponds to the Dirac
string. Thus, the Dirac strings have no action in the 
lattice, or compact version
of the $U(1)$ gauge theory. 

Note that the UV scale $1/a$
is the only scale in the model. Thus, it rather exists only as a lattice
theory. 
The confining potential sets in for $e>e_{crit}$ at all the distances
and in this sense the situation does not imitate QCD.

\section{Beyond the OPE: short strings.}

We can introduce another scale, apart from the UV cut off, by adding
a scalar field, i.e. by considering the Abelian Higgs model (AHM):
\beqn\label{AHM_action}
S= \int d^4x \big[
\frac{1}{4e^2} F^2_{\mu\nu} + \frac{1}{2} |(\partial - i A)
\Phi|^2 \\ \nonumber+ 
\frac{1}{4} \lambda (|\Phi|^2-\eta^2)^2
\big]
\eeqn
where $\lambda,\eta$ are constants.
The scalar field 
condenses in the vacuum, $\langle0|\Phi|0\rangle =\eta$
and this represents a new scale.
In particular, the physical 
vector and scalar particles are massive, $m^2_V=e^2\eta^2, m_H^2= 2
\lambda \eta^2$. Moreover, to avoid the monopole condensation
discussed in the preceding section we consider $e$ small enough 
to neglect the dynamical monopoles.

Consider now two external magnetic charges separated by a distance $r$
which are
brought into the vacuum of the Abelian Higgs model.
The problem is to find the potential energy $V(r)$
at the distances $r$ much smaller than $m_H^{-1}, m_V^{-1}$
\cite{gpz}.
The energy is determined in the classical approximation and,
at first sight, the problem is not much of a challenge.
The crucial point, however, that
the problem is not yet properly formulated.
Namely, one has to impose an extra 
boundary condition, that is vanishing of the scalar field
along a mathematically thin line connecting the magnetic charges.
This condition, a kind of topological one was formulated in its generality
in Ref. \cite{abelian} and implied in numerical studies of the problem
(see, e.g., \cite{alcock}). Moreover, the AHM is quite a common tool in
phenomenological studies of QCD at large distances (see, e.g., 
\cite{brambilla}). The {\it short-distance} aspects of the model have only
recently been emphasized \cite{gpz}.

The boundary condition $\Phi=0$ along a line
connecting the monopoles plays a crucial role.
It is worth emphasizing therefore that
 this infinitely thin line is nothing
else but the Dirac string connecting the external monopoles. 
As we discussed in the preceding section, 
the use of the lattice regularization implies a 
vanishing energy of the Dirac string
in the perturbative vacuum.
This is, in a way, a definition, how we understand the theory.
Moreover, it is easy to realize 
 that the Dirac
string cannot coexist with $\Phi\neq 0$. 
Indeed, if the Dirac string would be embedded into a vacuum with
$\langle \Phi\rangle\neq 0$ then its energy would 
again jump to infinity since there is the
term $1/2|\Phi|^2A_{\mu}^2$ in the action and $A_{\mu}^2\rightarrow
\infty$ for a Dirac string.  Hence, $\Phi=0$ along the string and it
is our boundary condition. In other words, Dirac
strings always rest on the perturbative vacuum which is defined as the
vacuum state obeying the duality principle. Therefore, even in the
limit $r\to 0$ there is a deep well in the profile of the Higgs field
$\Phi$. This might cost energy which is linear with $r$ even at small
$r$.

Thus, we are coming to 
the next question, whether this mathematically thin
line realizes as a short physical string. Where by the physical
string we understand a stringy piece in the potential, $\sigma_0\cdot r$
at small $r$.
In other words, we are going to see whether the
stringy boundary condition implies a stringy potential.
To get the answer one solves the classical equations
of motion. The answer is, indeed,
\be
\lim_{r\to 0}V(r)~=~-{\alpha_M\over r}+\sigma_0\cdot r
\ee
with a non-vanishing slope $\sigma_0$. It might worth emphasizing that
this result is actually non-trivial. Indeed, naively, one could
argue that the strong field near the monopoles ``burns the Higgs field
out'' anyhow and the condition $\Phi=0$ at small $r$ is not important for
this reason. In reality, however, this condition is manifested
in $\sigma_0\neq 0$. The form of the boundary condition finds its way
to the final result for the energy and this can be viewed
as a kind of analyticity. 

The slope $\sigma_0$ depends smoothly on the value of $m_V/m_H$.
For the purpose of orientation let us note that for $m_V=m_H$
the slope of the potential at $r\to 0$ is the same
as at $r\to \infty$ That is, within error bars:
\be
\sigma_0~\approx~\sigma_{\infty}\label{ssimple}
\ee
where $\sigma_{\infty}$ determines the value of the potential at large $r$.

The linear correction (\ref{ssimple}) to the potential
violates the OPE. It is not the same obvious as in case of QCD, though,
since now we do have an operator of $d=2$, that is $|\Phi|^2$.
However, a little more elaborated analysis than just counting the dimensions
still demonstrates the violation of the OPE \cite{gpz}.
Qualitatively, it is indeed quite obvious that the topological boundary 
condition cannot be reproduced via the OPE  but is imposed
extra.   

{\it To summarize}, existence of short strings has been proven 
in the classical approximation to
the Abelian Higgs model. 
The linear piece in the potential at small distances 
reflects the boundary condition that $\Phi=0$
along the straight line connecting the
monopoles and violates the OPE. This is a manifestation of
UV non-perturbative divergences in the energy of the Dirac string (see above).

\section{Monopoles in QCD.}

While monopoles are well known as classical solutions
in theories with Higgs fields, they can also be defined 
in a pure topological way \cite{abelian}. 
There is convincing evidence from the lattice simulations
that monopoles as defined in the maximal Abelian projection
condense in the QCD vacuum and dominate non-perturbative degrees of freedom,
see, e.g., \cite{polikarpov,digiacomo}. 

We refer the reader for any
detail on the QCD monopoles to \cite{polikarpov,digiacomo}
and references therein and address here
only one particular issue. Namely,
from the experience with the Dirac strings we learn that
pure topological defects in gauge theories are usually related
to some singular field configurations and, generally speaking, infinite
action. Only upon regularizing these UV non-perturbative
divergences to zero one allows these fluctuations into theory and
they may change the content of the theory completely.
What are singular fields associated with the monopoles, if any?
We would not be able to answer this question in full because of the lack of
understanding the large-distance behavior, or infrared physics which
is to provide a kind of mass for the charged gluons (as defined 
within the context of the Abelian projection \cite{abelian}).
However, an educated guess can be made about the small-distance
singularities.

The idea is to exploit the classical monopole solution.
As is well known the gauge field associated with the
monopole solution can be represented as
\be\label{potentials}
A^a_{\mu}~=~{\frac {f(r)}{g r^2}}\epsilon^{a\mu b}r^b,~\mu=1,2,3;~A_4^a=0.
\ee
The asymptotic behavior of the function $f(r)$ at small and large
distances is crucial:
\be
\lim_{r\to \infty}{f(r)}~=~1,~~~\lim_{r\to 0}f(r)~=~2~or~0\label{asymptotic}
\ee
(see, e.g., \cite{vinciarelli})
where by $r\to 0$ we still understand $r\gg m_H^{-1}$ 
since we are removing now the Higgs field altogether from
the theory, to imitate the QCD.
For the sake of estimates, 
the $r\to \infty$ asymptotic (\ref{asymptotic}) can be used down to distances
of order $m_V^{-1}$ which in the QCD set up is to be replaced
by $\La^{-1}$. On the other hand, the small $r$ asymptotic can also be
stretched to $r\sim m_V^{-1}$ from below. It is worth emphasizing that
the non-Abelian field strength tensor $G_{\mu\nu}^a\approx 0$ at small $r$.
As a result from the Eq. (\ref{uvdivergence}) we would obtain
estimate the integral
$\int d^3r (G^a_{\mu\nu})^2$ as $m^{-1}_V/g^2$ which is correct.

Now start first with large distances and perform the gauge transformation
which brings (\ref{potentials}) to the Abelian form 
(see, e g., \cite{polikarpov}). Then at small
distances we generate a Dirac string which is directed along the
third axis in the color space and along the $x_3$-axis in the space:
\beaq{l}
G_{\mu\nu}=
-
\sigma^3  \; 
( \delta_{\mu,1}\delta_{\nu,2} -\delta_{\mu,2}\delta_{\nu,1}) \cdot
\\
\rule{30mm}{0mm}\rule{0mm}{5mm}%
\cdot \; \frac{\D{2\pi}}{\D{g}} \;
\delta(x_2)\delta(x_1)\Theta(-x_3)
\label{abelianstring} \eeaq
Moreover the corresponding potential is singular at small $r$.
For example, in the spherical coordinates, the $\theta$-component
of the potential is now given by:
\be
A_{\theta}^{1+i2}~=~{e^{\i\phi}\over r}.\label{theta}
\ee
Also, if we start directly with the small $r$ asymptotic 
$\lim_{r\to 0}{f(r)}=2$
we would arrive at a potential which is singular at $r=0$:
\be
A_{\mu}^a~=~-i \; \mbox{Tr}\left[
\sigma^a \;
(\vec{\sigma}\cdot\vec{n}) \;
\partial_{\mu} \;
(\vec{\sigma}\cdot\vec{n})
\right]
\label{singular}
\ee
where $\vec{n}$ is the unit vector directed from the center of the
monopole to the observation point and $\vec{\sigma}$ are the 
Pauli matrices.

If we take the potential (\ref{singular})
at face value, then we would conclude that
the associated action is divergent in ultraviolet and infinite.
Indeed, the potential (\ref{singular}) has a $\delta(r)$ type
of singularity at $r=0$. Thus, naively, the action is violently divergent:
$$\int ((G_{\mu\nu}^a)^2d^3x~\sim\delta (r)$$
If there is a Higgs field then all the fields are smoothened at $r=0$.
Moreover, the contribution of the region $r\sim m_H^{-1}$ to the total
action of the monopole solution is negligible. In this sense, the only
function of the Higgs mass is to provide an ultraviolet regularization
which allows for singular potentials (\ref{theta}),(\ref{singular}).

Now we come to the 
central point of UV regularization of the singular potentials
in QCD. Since there is no Higgs field at all we should apply another
regularization. The natural choice is of course the lattice regularization.
Then we know already from the discussion of the compact $U(1)$ above that
the Dirac string (\ref{abelianstring}) is associated with no action. 
A new point which is specific for the monopoles is that 
the monopoles are placed into centers of the lattice cubes and,
therefore, the lattice regularizes the $\delta (r)$ type of
singularities (see above) to zero as well\footnote{
The detailed consideration will appear soon elsewhere.
}.
Thus the 
singular potential (\ref{singular}) is allowed 
on the lattice and brings no action, the same as for the classical
solutions.

{\it To summarize}, we expect that the QCD monopoles are associated with
potentials which are singular at the origin. The lattice 
UV regularization
implies that such potential are allowed, however. 
A qualitative picture for the QCD monopoles would be that they
are associated with singular potentials which bring
no action at small $r$. Overlapping potentials of this type,
however, are no longer solutions to $G^a_{\mu\nu}=0$
and there is  suppression because of the action associated with
such an overlap. However, 
there is a gain in the entropy as well
and one can expect that the balance is reached at $g^2\sim 1$ since the
entropy factors have no small parameter built in (see the discussion
of the compact $U(1)$ above).
Of course, our consideration 
is absolutely
incomplete as far as 
there is no understanding of the physics at $r\sim \La^{-1}$.
In particular, as far as the $r\sim 0$ region is considered in isolation,
the potential can be gauge rotated to zero.

\section{A remark on the P-vortices.}
Amusingly enough, the
qualitative picture outlined in the conclusions
to the preceding subsection might be subject to 
an experimental check by means of the P-vortices.
While referring the reader for any detail about the P-vortices
to a few recent articles
\cite{bertle} and references therein,
here we just mention a few basic facts about the P-vortices.

One usually uses a specific gauge, namely,  the gauge 
maximizing the sum
\be
\sum_{l}{|Tr~U_{l}|^2}
\ee
where $l$ runs over all the links on the lattice.
The center projection is obtained by replacing
\be
U_{l}~\rightarrow~sign~ (Tr~U_{l}).
\ee
Each plaquette is marked either as (+1) or (-1)
depending on the product of the signs assigned to the 
corresponding links.
P-vortex then pierces a plaquette with (-1). 
Moreover, the fraction $p$ of the total number of plaquettes pierced by the
P-vortices and of the total number of all the plaquettes $N_T$,
obeys the scaling law
\be
p~=~{N_{vor}\over N_T}~\sim~f(\beta )
\ee
where the function $f(\beta)$ is such that
$p$ scales like the string tension.
Assuming independence of the piercing for each plaquette
one has then for the center-projected Wilson loop $W_{cp}$:
\be
W_{cp}=[(1-p)(+1)+p(-1)]^A~\approx~e^{-2pA}\label{tension}
\ee
where $A$ is the number of plaquettes in the area stretched on
the Wilson loop. 
Numerically, Eq. (\ref{tension}) reproduces the full string tension.

Now we argue that P-vortices correspond to large gauge potentials
$A_\mu^a$. Really the statement ``large (or small) gauge potential''
is obviously gauge dependent, and below we will discuss the maximal
central gauge in which P-vortices are usually defined. Suppose that the
plaquette is pierced by P-vortex, that is one or three links
forming this plaquette have negative trace ($Tr \, U_l < 0$). 
It is easy to
show that the links with negative trace correspond to large (infinitely
large in the continuum limit) gauge potential.

Consider the link matrix, which correspond to the gauge potential
$A_l^a$:
\be
U_l = \exp\{i\frac{\sigma^a}{2} A_l^a\} =
\cos\frac\theta 2 + i\vec{\sigma}\vec{n} \, sin \frac\theta 2
\ee
where $\theta = \sqrt{\sum_a |A_l^a|^2}$, and $n^a = A_l^a/\theta$ is
the unit vector.
If $Tr \, U_l < 0$ then $\cos\frac\theta 2 < 0$ or
$\theta > \pi$. 
Recovering dimensional quantities
we have:
$\sqrt{\sum_a |A_l^a|^2} > \frac \pi a$,
where $a$ is the lattice spacing. Hence if P-vortices
survive in the continuum then it means that in the continuum limit
we have gauge potentials $A_l^a$ of the order of the cutoff
$\frac 1 a$. 
Thus P-vortices correspond to large gauge potentials just as field 
configuration (\ref{singular}), discussed in the previous section.
Remarkable fact is that there exist a strong correlation of P-vortices and
abelian monopoles, although these objects are defined in different 
gauges \cite{data}.

It is important that the P-vortex is constructed
in a gauge-dependent way which makes it sensitive  to large
(i.e., non-accessible perturbatively) fields,
but not large field strengths. This statement follows simply
from the fact that P-vortices survive in the continuum limit.
Let us note that practically all the papers which
discuss P-vortices in the continuum
(see \cite{casimir2,reinhardt}
and references therein) use the formalism which corresponds to
the assumption that the P-vortices pierce in fact plaquettes
characterized by a negative sign of the Wilson loop. 
The idea that the P-vortex is sensitive to strong potentials rather
than field strengths promoted in this talk has not been tested,
to the best of our knowledge.

In the particular gauge P-vortices appear to be 
infinitely thin. Hence
they  prove in a sence that UV 
non-perturbative physics can be related
to the physics of confinement. Note 
that P-vortices are not truly local objects:
there definition involves the nonlocal gauge 
fixing procedure, in any other gauge 
P-vortex will be clearly nonlocal field
configuration (for analogous 
discussion of gauge dependent operators
in case of abelian projection see  \cite{yukis97}).
However, this relevance of the UV non-perturbative fields could well be
an artefact of the gauge used. Violation of the OPE,
on the other hand, would be a gauge
independent proof of the relevance of the strong fields,
as is explained in length in preceding subsections.
From this point of view, the linear correction
to the heavy quark potential would be most important. 
And, so far, the linear non-perturbative potential due to
the P-vortices is observed numerically
down to distances equal to a single
lattice spacing.

In other words, Eq. (\ref{tension}) appears to work for all the distances.
The violation of the standard picture is then traced to the fact
that the P-vortices are infinitely thin.
Indeed, what is crucial for the standard picture is a finite size
of the non-perturbative fluctuations \cite{akz}.
Note that in Ref \cite{casimir2} an attempt was made to modify (\ref{tension})
by assuming that the piercing of the plaquettes by the P-vortex is
not random at distances below the thickness of the center vortex
associated
with the P-vortex. The resulting heavy quark potential immediately
becomes quadratic at short distances, in accordance with the general
theory. Thus, measurements of the potential seem to be of crucial 
importance.

{\it To summarize}, we argued that the P-vortices
can be interpreted as evidence for essential role
of the potentials which
are singular in the continuum limit
and survive on the lattice because of the lattice regularization. 
Measurements which demonstrate a linear potential due to the P-vortices
at short distances indicate then that this role is not a mere artifact
of the gauge used. In this sense confirmation
of the existence of the linear correction would be of great importance.

\section{Revisiting phenomenology.}

In this section we come back to discuss (preliminary) evidence
contradicting the standard picture and try to understand whether such
evidence could be accommodated into the picture ``beyond the OPE''.
It is worth emphasizing from the beginning that the breaking of the
OPE was demonstrated theoretically only within the AHM \cite{gpz}.
The consideration of the QCD case outlined above 
by itself, i.e. without input of the data, is inconclusive.
Thus, it is clear that at this moment phenomenology of the novel type 
power corrections can be based only on extra assumptions. It is
amusing, nevertheless how well some simple assumptions work and
we review briefly these assumptions.

{\it Heavy quark potential at short distances}.

It is well known that the infrared behavior of QCD is well described by
the Abelian Higgs model with $m_V\approx m_H$, see, e.g., \cite{suzuki}. 
By assuming the analogy to be valid for the vacuum state
also at short distances, one would predict $\sigma_{\infty}\approx\sigma_0$
(see Eq. (\ref{ssimple}). This prediction fits surprising well
the potential induced by the P-vortices (see above).

{\it Shifts of the $Q\bar{Q}$ atomic levels.}

If, indeed, $\sigma_0\approx \sigma_{\infty}$ then
the predictions for the energy levels would be close to those obtained 
within the phenomenological Buchmuller-Tye potential \cite{buchmuller}
which incorporates $\sigma_0\approx\sigma_{\infty}$:
\be
(\delta E)_{non-pert}~\approx~(\delta E)_{Buchmuller-Tye}   
.\ee
These predictions seem to be in much better shape in view of the lattice
data \cite{fingberg} than the standard picture.

{\it Correlation functions.}

If the OPE is violated, there could arise corrections of order
$1/Q^2$ in the r.h.s. of the Eq. (\ref{normal}):
\beqn
\Pi(Q^2)\approx \Pi_j(Q^2)_ {parton~model}\cdot\\
\big(1+{a_j\over \ln[Q^2/\Lambda^2_{QCD}]}+{b_j\over Q^4}+{c_j\over Q^2} + ...\big)
\eeqn
Even if one accepts this assumption, it is far from being trivial
to relate the new constants $c_j$ to, say,
linear term in the potential.  

Qualitatively, however, one may hope that
introduction of a tachyonic gluon mass at short distances  
would imitate the effect of the 
$\LQ^2/Q^2$ corrections. Indeed, the stringy term in
$V(r)$ at short distances
can be imitated \cite{vz2} by the Yukawa 
potential with a gluon mass $\lambda $:
\be
{4\alpha_s\over 6}\lambda^2~\sim~-~\sigma_0\label{mass}
.\ee
While Eq (\ref{mass}) by itself 
is nothing else but a way to memorize the result
for the potential at short distances, it is acquiring predictive power once
one introduces the short distance tachyonic mass into one-loop graphs
as well \cite{cnz}. Amusingly enough, this bald 
assumption allows to explain paradoxes of the standard picture (see above)
in a simple and unified way. 

To begin with, phenomenologically,
in the $\rho$-channel there are severe restrictions \cite{narison}
on the new term $c_j/Q^2$:
\be
c_{\rho}~\approx~-~(0.03-.07)~GeV^2\label{constr}
.\ee
Remarkably enough, the sign of $c_{\rho}$
does correspond to a tachyonic gluon mass
(if we interpret $c_{\rho}$ this way).
Moreover, when interpreted in terms of $\lambda^2$ the constraint (\ref
{constr}) does allow for a large $\lambda^2$, say, $\lambda^2=-0.5GeV^2$.

As for for the $\pi$-channel one finds now a new value of $M^2_{crit}$
associated with $\lambda^2\neq 0$:
\be
M^2_{crit}(\pi-channel)\approx~4\cdot M^2_{crit}(\rho-channel)\ee
which fits nicely the Eqs.
(13) and (14) above. 
Moreover, the sign of the correction in the $\pi$-channel 
is what is needed for phenomenology \cite{novikov}.
Fixing the value of $c_{\pi}$ to bring the theoretical $\Pi_{\pi}(Q^2)$
into agreement with the phenomenological input one gets
\be
\lambda^2~\approx~-0.5~ GeV^2
.\ee
Finally, we can determine the new value of $M_{crit}^2$ in the 
scalar-gluonium channel and it turns to be what is 
needed for the phenomenology,
see Eq (15).

Further crucial tests of the model with the tachyonic gluon mass could be 
furnished with measurements of various correlators $\Pi_j(Q^2)$
on the lattice \cite{cnz}. 

{\it To summarize,}in spite of the openly heuristic nature,
the model with a short-distance tachyonic gluon mass works surprisingly
well. Indeed, it resolves the long-standing paradoxes of the QCD sum rules
without spoiling the successful predictions.
At this moment, this model seems to be the best candidate for
a ``dogma'', although this may change any time with arrival of new data.
Let us also mention some particular dynamical schemes \cite{simonov}
which come close to imitation of the tachyonic gluon mass.

\section{Conclusions}

We have argued that a healthy phenomenology of the power corrections
going beyond the OPE can be developed at this moment \cite{cnz,gpz}.
This observation by itself comes as a kind of surprise since the power
corrections have been studied for more than 20 years. 
On the theoretical side, the violation of the OPE has been proven in
case of the Abelian Higgs model \cite{gpz}. In case of QCD, the results
of the analysis outlined above are inconclusive.

\end{document}